\documentclass[twocolumn]{revtex4}
\usepackage[dvips]{graphicx}
\begin{document}

\title{
Metal-insulator transition in Ca$_{1-x}$Li$_x$Pd$_3$O$_4$ 
}

\author{S. Ichikawa$^1$ and I. Terasaki$^{1,2}$}
\email{terra@waseda.jp}
\affiliation{
$^1$Department of Applied Physics, Waseda University, Tokyo 169-8555, Japan\\
$^2$Precursory Embryonic Science and Technology, 
Japan Science and Technology Corporation, Tokyo 108-0075, Japan
}

\date{\today}

\begin{abstract}
Metal-insulator transition in Ca$_{1-x}$Li$_x$Pd$_3$O$_4$
has been studied through charge transport measurements.
The resistivity, the Seebeck coefficient, and the Hall coefficient
are consistently explained in terms of a simple one-band picture,
where a hole with a moderately enhanced mass 
is itinerant three-dimensionally.
Contrary to the theoretical preditcion 
[Phys. Rev. B62, 13426 (2000)], 
CaPd$_3$O$_4$ is unlikely to be an excitonic insulator, and
holds a finite carrier concentration down to 4.2 K.
Thus the metal-insulator transition
in this system is basically driven by localization effects.
\end{abstract}

\pacs{}

\maketitle

\section{Introduction}
A metal-insulator transition is one of the most important topics 
in solid-state physics \cite{mott}.
In case of a conventional semiconductor, 
donors (acceptors) supply  electrons (holes) in a band insulator.
A metal-insulator transition takes place at a critical carrier density,
where the average carrier-carrier distance is comparable 
with the effective Bohr radius of the doped impurity atom.
The critical carrier density has been quantitatively discussed
for various semiconductors \cite{belitz}.

Metal-insulator transitions in strongly correlated systems 
are completely different from that in a band insulator \cite{tokura}.
The most remarkable example is that in  high-temperature superconductors,
where the metallic state is accompanied by high temperature superconductivity.
As the second example, the colossal magnetoresistive manganites
exhibit a peculiar transition where the metallic state
is stabilized by ferromagnetism.
Actually, there are many types of insulating state
(Mott insulator, charge ordering, and stripes)
for strongly correlated systems, 
and accordingly there are so many ways how these insulating states
collapse upon doping, pressure, temperature, and external field.

Since transition-metal oxides are often insulating due to the 
strong correlation, they can be a good playground for 
studies of metal-insulator transitions.
In this sense, we have paid special attention to CaPd$_3$O$_4$.
Figure 1 shows the crystal structure of CaPd$_3$O$_4$ known 
as a NaPt$_3$O$_4$-type structure.
The divalent Pd$^{2+}$ of (4$d$)$^{8}$ is surrounded with four O$^{2-}$ anions,
and the PdO$_4$ clusters stack one another to form a column.
Because of the cubic symmetry, the PdO$_4$ column runs along 
the $x$, $y$, $z$ directions to make a thee-dimensional network.

\begin{figure}[b]
 \begin{center}
  \includegraphics[width=6cm,clip]{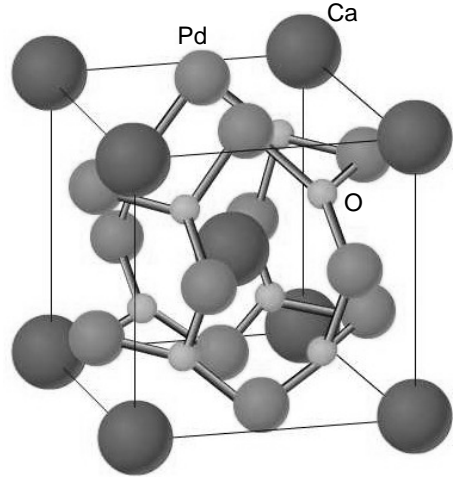}
 \end{center} 
 \caption{
 Crystal structure of CaPd$_3$O$_4$.
 }\label{f1}
\end{figure}

The first interesting point is that CaPd$_3$O$_4$ 
shows a metal-insulator transition by substitution of Na for Ca, 
as was found by Itoh et al \cite{itoh, itoh2}.
Secondly, the ground state of CaPd$_3$O$_4$ might be exotic.
Hase and Nishihra \cite{hase} claimed that CaPd$_3$O$_4$ was 
a possible candidate for an excitonic insulator, 
in which electrons and holes bounded as excitons
exhibit a Bose-Einstein condensation at low temperatures.
Thirdly, a Pd ion is more stable as Pd$^{2+}$ and Pd$^{4+}$ than Pd$^{3+}$.
Such a kind of ion is called ``valence skipper''.
Thus the doped holes are most likely to exist as Pd$^{4+}$,
where two holes are on the same Pd site.
Varma \cite{varma} predicted that doped carriers in the valence skipper 
form on-site pairs to exhibit a possible high-temperature superconductivity.

In this paper we report on measurement and analysis of the transport 
properties of Li doped CaPd$_3$O$_4$.
We have found that CaPd$_3$O$_4$ is essentially a degenerate semiconductor 
of low carrier concentration (10$^{19}$ cm$^{-3}$).
With increasing Li content, the resistivity, the Seebeck coefficient, 
and the Hall coefficient systematically change,
from which the carrier concentration and the effective mass 
are reasonably evaluated.

\section{Experimental}
Polycrystalline samples of Ca$_{1-x}$Li$_x$Pd$_3$O$_4$
($x$=0, 0.1, 0.2, 0.3, 0.4, 0.5 and 0.6)
were prepared by a solid-state reaction aided with NaCl addition.
Stoichiometric amount of PdO, CaCO$_3$, and LiCO$_3$ of 99.9\% purity
were thoroughly mixed.
NaCl was then added with the mixture in a mass ratio of 2:1, 
thoroughly mixed again, and fired at 1073 K for 24 h in air.
The product was finely ground, and NaCl was rinsed out in hot
distilled water.
The dried powder was then pressed into pellets, and was sintered
in air at 1223 K for 48 h for $x$=0, and at 1103 K for 12 h for $x>0$ .

We think that NaCl acts as a kind of flux. 
At an early stage of this study, 
20\% of metal Pd appeared as an impurity phase 
in the powder sintered above 1173 K without NaCl,
suggesting that the sintering temperature was too high to keep Pd oxidized.
We then decreased the sintering temperature, but found that the prepared 
samples were loosely sintered, too  fragile to handle.
Added NaCl was effective to decrease the sintering temperature 
without reducing PdO.

\begin{figure}[t]
 \begin{center}
  \includegraphics[width=7cm,clip]{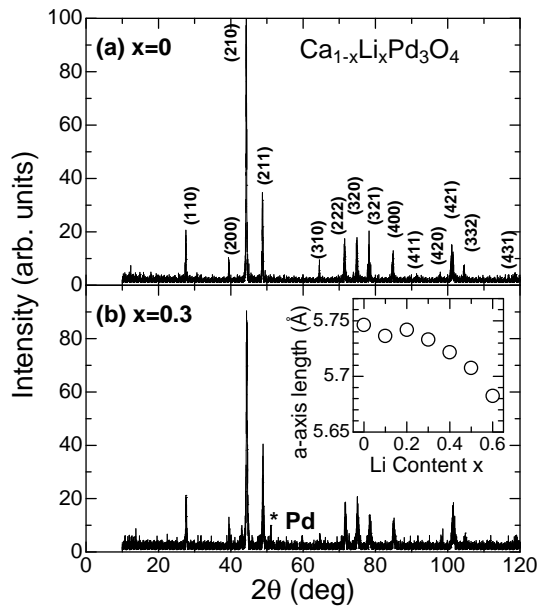}
 \end{center} 
 \caption{
 X-ray diffraction pattern  of CaPd$_3$O$_4$.
(a) $x=0$, and (b) $x=0.3$
 }\label{f2}
\end{figure}

The samples were characterized through the x-ray diffraction (XRD)
with Fe K$_{\alpha}$ as an x-ray source in a $\theta-2\theta$ scan mode.
The resistivity was measured by a four-terminal method from 
4.2 to 300 K in a liquid He cryostat,
and from 300 to 700 K in a cylinder furnace in air.
The Seebeck coefficient was measured using a steady-state technique 
with a typical temperature gradient of 1.0 K/cm 
from 4.2 to 300 K in a liquid He cryostat,
and from 300 to 500 K in a cylinder furnace in air.
The Seebeck coefficient of the voltage leads was carefully subtracted.
The Hall coefficient was measured in a closed refrigerator
from 10 to 150 K. 
A cernox resistive thermometer was placed at 45 cm above 
the magnet core, which successfully suppressed the 
magnetoresistance of the thermometer to keep the accuracy of
the measured temperature within 0.01\% at 7 T.
An ac-bridge nano-ohmmeter was used to measure the resistivity 
by sweeping magnetic field from -7 to 7 T in 20 minutes 
at constant temperatures.
An unwanted signal occurring from a misalignment of the 
voltage pads was carefully removed by subtracting 
negative-field data from positive-field data.
The Hall voltage was linear in magnetic field,
and the Hall coefficient was determined by the data at $\pm$7 T.

\section{Results and discussion}
Figure 2 shows typical XRD patterns of the prepared samples.
For $x=0$, all the reflection peaks are indexed 
as a NaPt$_3$O$_4$-type structure
with an a-axis length $a_0$ of 5.74 \AA \cite{wnuk}.
As is shown in the inset, $a_0$ decreases systematically with $x$,
which clearly indicates that Li$^+$ (0.76\AA) and Ca$^{2+}$ (1.00\AA) 
make a solid solution.
For $x=0.3$, however, a tiny (5\%) trace of reduced Pd is observed 
near $2\theta=$51 deg. 
The amount of the reduced Pd gradually increases with $x$ for $x>0.3$,
and exceeds 10\% for $x>0.6$.

\begin{figure}[t]
 \begin{center}
  \includegraphics[width=6.5cm,clip]{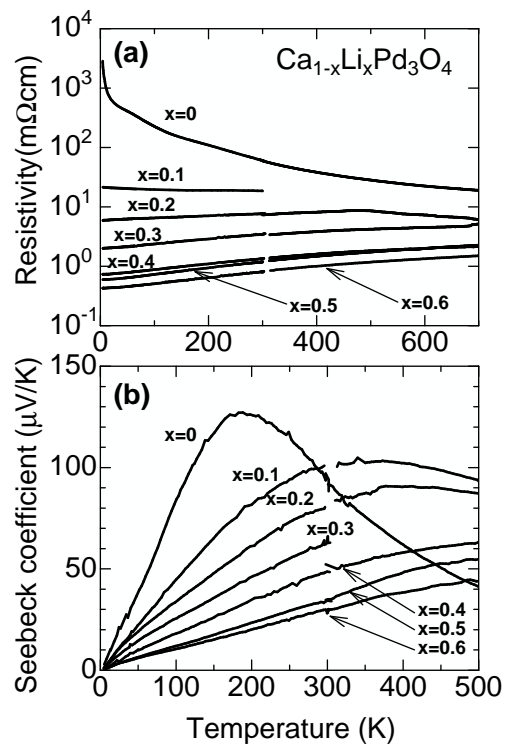}
 \end{center} 
 \caption{
 (a) Resistivity, and (b) the Seebeck coefficient
 of Ca$_{1-x}$Li$_x$Pd$_3$O$_4$
 }\label{f3}
\end{figure}

Figure 3(a) shows the temperature dependence of the resistivity 
($\rho$) of the prepared samples.
The resistivity changes systematically with $x$,
whose magnitude at 4.2 K decreases by four orders of 
magnitude from $x=0$ to 0.6.
This clearly indicates that the substituted Li supplies carriers
into the sample.
Towards 0 K,  $\rho$ for $x=0$ divergingly increases
while $\rho$ for $x=0.1$ remains a finite value.
This means the metal-insulator transition takes place between
$x=0$ and 0.1.

 Figure 3(b) shows the temperature dependence of the Seebeck coefficient
($S$) of the prepared samples.
All the Seebeck coefficients are positive, 
indicating that the majority carrier is a hole in this system.
$S$ for $x=0$ is roughly proportional to temperature below 100 K, 
which is a hallmark of a conventional metal.
In the sense that a finite carrier concentration 
remains as $T \to$ 0, CaPd$_3$O$_4$ is essentially a metal.
Considering the large thermopower, it would be more
appropreate to regard CaPd$_3$O$_4$ as a degenerate semiconductor
rather than a metal.
With increasing $x$, $S$ systematically decreases, 
which indicates that the substituted Li
supplies holes to Ca$_{1-x}$Li$_x$Pd$_3$O$_4$.
The doped samples also show a similar $T$-linear Seebeck coefficient
at low temperatures to the $x=0$ sample.
At high temperatures for smaller $x$, the Seebeck coefficient 
takes a maximum and decreases with increasing temperature.
This means that the minority carriers (electrons in this case)
are thermally excited,
and also means that the band gap of CaPd$_3$O$_4$ is smaller than 
hundreds of K.

Figure 4 shows the temperature dependence of the Hall coefficient 
($R_H$) of the prepared samples.
As is similar to the Seebeck coefficient, the sign of the Hall coefficient
is positive for all the samples, indicating the majority carriers
are holes.
The magnitude systematically decreases with increasing $x$, and
$R_H$'s for $x=0.1$ and 0.2 are essentially independent of temperature.
These results are what is expected in a conventional metal, 
magnitude of which is inversely proportional to the carrier concentration.

\begin{figure}[t]
 \begin{center}
  \includegraphics[width=6cm,clip]{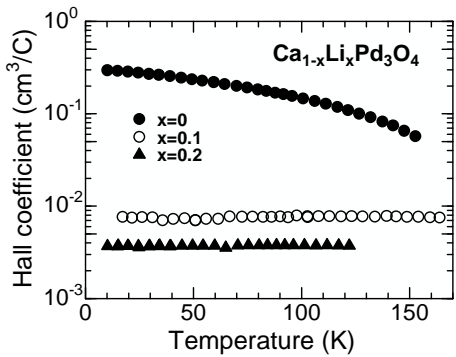}
 \end{center} 
 \caption{
 The Hall coefficient
 of Ca$_{1-x}$Li$_x$Pd$_3$O$_4$
 }\label{f4}
\end{figure}

We should note that $R_H$ for $x=0$ saturates as $T\to 0$.
This clearly indicates that CaPd$_3$O$_4$ is essentially 
a degenerate semiconductor 
with a low carrier concentration of 10$^{19}$ cm$^{-3}$,
which is consistent with the $T$-linear $S$.
Contrary to the theoretical prediction by Hase and Nishihara \cite{hase},
CaPd$_3$O$_4$ is unlikely to be an excitonic insulator 
having the charge gap of the order of an exciton 
binding energy at low temperatures.
Thus we conclude that the nonmetallic $\rho$ for $x=0$ 
is due to localization effects.
$R_H$ for $x=0$ exhibits remarkable temperature dependence, 
which implies the existence
of the additional carriers activated thermally.
This further suggests that the activation energy is smaller than
a few tens of K, because $R_H$ is clearly 
dependent on temperature down to 10 K.
The small activation energy can also explain the 170-K peak in $S$ 
for $x=0$,
where the thermally activated electrons dominates at high temperatures.
It is not surprising that $R_H$ and $S$ show different temperature dependences.
According to a two-band model consisting of electrons and holes, 
$R_H$ and $S$ are averaged with different weights of 
electron ($\sigma_e$) and hole ($\sigma_h$) conductivities.
A more quantitive analysis would be difficult,
unless $\sigma_e$ and $\sigma_h$ were experimentally determined.

A possible candidate for the small activation energy is the band gap 
of CaPd$_3$O$_4$. 
According to the intuitive explanation by Doublet et al. \cite{doublet}, 
the valence band of CaPd$_3$O$_4$ consists of Pd 4$d_{z^2}$, 
while the conduction band consists of Pd 4$d_{x^2-y^2}$ hybridized strongly
with O 2$p$.
Owing to the large dispersion of the Pd 4$d_{z^2}$ band,
the energy gap between the valence and conduction bands
is expected to be small.
In fact, the band calculation \cite{hase} 
showed that CaPd$_3$O$_4$ was a semi-metal with zero band-gap.

Now we will make a quantitative discussion on the metal-insulator 
transition in CaPd$_3$O$_4$.
As already mentioned, Li substitution supplies holes in CaPd$_3$O$_4$,
and the doped holes seem ``normal'', in the sense that 
they give the $T$-linear Seebeck coefficients
and the $T$-independent Hall coefficients at low temperature.
Thus we employ the simplest formula of $R_H=1/{ne}$, 
where $n$ is the hole concentration. 
Then $R_H$'s for $x=0.1$ (7.7$\times$10$^{-2}$ cm$^3$/C)
and $x=0.2$ (3.7$\times$10$^{-2}$ cm$^3$/C)
correspond to 0.08 and 0.16 holes per unit formula, respectively.
In spite of the rough estimation, the hole concentration obtained from $R_H$ 
is in good agreement with $x$,
and we can roughly assume that the Li content $x$ 
supplies $x$ holes per Li cation.

\begin{figure}[t]
 \begin{center}
  \includegraphics[width=8cm,clip]{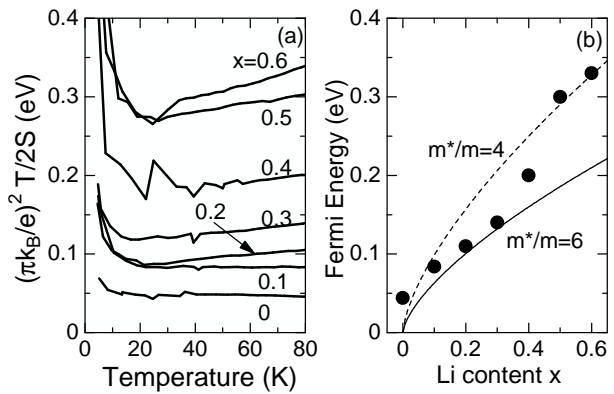}
 \end{center} 
 \caption{
 (a) The Fermi energy evaluated from the Seebeck coefficient.
 (b) The Fermi energy plotted as a function of Li content.
 Dotted and solid curves represent the Fermi energy 
 calculated from a nearly free hole picture with an
 effective mass of 4 and 6 (see text).
 }\label{f5}
\end{figure}

In the lowest order approximation, the Seebeck coefficient of
a conventional metal is expressed as 
$$
S=-\frac{\pi^2k_B^2}{2eE_F}T 
$$
where $E_F$ is the Fermi energy. 
Then $E_F$ can be experimentally determined from the $T$-linear part
of the Seebeck coefficient as $E_F=\pi^2 k_B^2T/2e^2S$ [eV].
Figure 5(a) shows $\pi^2 k_B^2 T/2e^2S$ for the prepared samples.
All the data are roughly independent of temperature at 80 K,
from which we evaluate $E_F$ for each sample.
We think that $T/S$ at low temperatures is less reliable, 
owing to the small magnitude of $S$
and low sensitivity of the copper-constantan thermocouple below 10 K.

Figure 5(b) shows the evaluated $E_F$ plotted as a function of $x$.
We further assume a nearly free hole with an effective mass of $m^*$,
the Fermi energy is written as 
$$ 
E_F=\frac{\hbar^2}{2m^*}\left( 3\pi^2n \right)^{\frac{2}{3}}
   =\frac{\hbar^2}{2m}\left(3\pi^2\frac{2x}{a_0^3} \right)^{\frac{2}{3}}
   \frac{m}{m^*}.
$$
where $m$ is the bare mass of an electron (Note that a unit cell 
includes two unit formulae).
The dotted and solid curves show the calculated $E_F$ with $m^*/m=$4 and 6,
respectively, between which the evaluated $E_F$ lies.
This means that the effective mass of Ca$_{1-x}$Li$_x$Pd$_3$O$_4$
is nearly independent of $x$, and is moderately (4-6 times)
enhanced from the bare mass,
possibly owing to the 4d nature.

Finally we will make brief comments on remaining issues.
(i) Ca$_{1-x}$Li$_x$Pd$_3$O$_4$ is a possible candidate 
for a p-type thermoelectric oxide \cite{mahan}.
A thermoelectric material is a material that converts heat into 
electric power, and electric power into heat, through 
the thermoelectric phenomena in solids.
The thermoelectric power factor $S^2/\rho$ is 1.6 $\mu$W/cm K$^2$
for $x=0.4$ at 300 K, which is comparable to the value
for the polycrystalline NaCo$_2$O$_4$ known as a promising candidate 
for a thermoelectric oxide \cite{terra}.
(ii) At present, we have no direct evidence that
the doped holes form on-site pairs like Pd$^{4+}$.
The charge transport observed in the present paper
is quantitatively explained in terms of nearly free holes
with the enhanced mass.
In other words, this system is highly robust against 
charge disproportionaltion and/or charge density wave.
The orthogonally entangled PdO$_4$ columns may play an important role, 
as was suggested by Doublet et al.\cite{doublet}
Another reason would be that the carrier concentration
was too low to observe the ``valence skipper'' effects.
In the case of Bi oxides, the valence skipper effects
are most remarkable in BaBiO$_3$, where the charge ordered
state of Bi$^{3+}$ and Bi$^{5+}$ are stabilized.
With doping, the Bi$^{3+}$-Bi$^{5+}$ state collapses
and the band picture gradually recovers.
In the present study, the Pd$^{4+}$ content is less than 10 \%, 
and this corresponds to BaPb$_{1-x}$Bi$_{x}$O$_3$ ($x<0.2$) 
which can be explained by the band picture \cite{uchida}.
Thus it would be more tempting to synthesize a Pd oxide
consisting of (formally) Pd$^{3+}$ to search for 
the valence skipper effects.
(iii) The small band gap implies that electrons can be doped in the 
conduction band.
Preliminarily we succeeded in electron doping by substitution of 
a trivalent ion (La$^{3+}$, Y$^{3+}$, and Bi$^{3+}$) for Ca$^{2+}$.
(iv) The present results are quite different from $\rho$ and $S$
for Ca$_{1-x}$Na$_x$Pd$_3$O$_4$ by Itoh et al.\cite{itoh, itoh2}
We prepared Na substituted samples, but found that $\rho$ and $S$ 
were essentially the same as those for Ca$_{1-x}$Li$_x$Pd$_3$O$_4$.
We employed ``NaCl flux'' technique to suppress reduction of PdO,
but Itoh et al. used conventional solid state reaction.
Thus the different preparation method might give
different samples.

\section{Summary}
The resistivity, the Seebeck coefficient, and the Hall coefficient
for Ca$_{1-x}$Li$_x$Pd$_3$O$_4$ ($x$=0-0.6)
have been measured and analyzed.
Since the parent material CaPd$_3$O$_4$ is a degenerate semiconductor
 with a finite carrier concentration of 10$^{19}$ cm$^{-3}$, it is unlikely to be
an excitonic insulator as suggested by Hase and Nishihara \cite{hase}.
The metal-insulator transition in this system is thus 
basically driven by localization effects.
The carrier concentration dependence of the Hall and Seebeck
coefficients is consistently explained in terms of a simple one-band picture,
where a hole with a moderately enhanced mass ($m^*/m\sim$4-6)
is itinerant three-dimensionally.

The authors would like to thank T. Fujii, R. Kitawaki, W.Kobayashi, 
and K. Kurihara for useful comments.
They also appreciate T. Sugaya for technical support.

\end{document}